# Search Driven Analysis of Heterogeneous XML Data

Andrey Balmin
IBM Almaden Research
San Jose CA, USA
abalmin@us.ibm.com

Latha Colby
IBM Almaden Research
San Jose CA, USA
lathac@us.ibm.com

Emiran Curtmola[*]
UC San Diego
San Diego CA, USA
ecurtmola@cs.ucsd.edu

Quanzhong Li
IBM Almaden Research
San Jose CA, USA
quanzhli@us.ibm.com

Fatma Özcan
IBM Almaden Research
San Jose CA, USA
fozcan@almaden.ibm.com

## ABSTRACT

Analytical processing on XML repositories is usually enabled by designing complex data transformations that shred the documents into a common data warehousing schema. This can be very time-consuming and costly, especially if the underlying XML data has a lot of variety in structure, and only a subset of attributes constitutes meaningful dimensions and facts. Today, there is no tool to explore an XML data set, discover interesting attributes, dimensions and facts, and rapidly prototype an OLAP solution.

In this paper, we propose a system, called SEDA[1], that enables users to start with simple keyword-style querying, and interactively refine the query based on result summaries. SEDA then maps query results onto a set of known, or newly created, facts and dimensions, and derives a star schema and its instantiation to be fed into an off-the-shelf OLAP tool, for further analysis.

## 1. INTRODUCTION

As XML repositories become pervasive, there is a pressing need to leverage this data in business intelligence applications. Due to the lack of native XML tools for on-line analytical processing (OLAP), users typically employ relational OLAP tools if they want to run analytics over an XML repository. This requires designing a warehouse schema, and creating complex data transformations that shred XML documents into this common schema. However, these tasks are known to be very time-consuming and costly, and most importantly, warehouse designers must be knowledgable about the details of the underlying XML data structures. On the other hand, in practice, XML repositories typically contain data with complex and varying formats. For example, in several widely used industry-standard XML schemas, such as XBRL (http://www.xbrl.org) and HL7 (http://www.hl7.org), there is a common schema, but it is very generic and contains many optional elements. As a result, the XML data conforming to these schemas have a lot of variety in their structure. Moreover, only a subset of the attributes in these schemas constitutes dimensions and facts that are meaningful to a particular analytical task. Today, there is no tool to explore an XML data set, discover interesting attributes, dimensions and facts, and rapidly prototype an OLAP solution.

As a first step to prototype an OLAP solution, the users need to find fragments of XML schemas and documents that are relevant to the analytical task at hand. Querying XML collections that contain lots of variety using schema-aware languages like XQuery[25] or XPath is cumbersome, and not always viable because XPath expressions are designed for structural navigation of XML data, and require a high-degree of schema knowledge. On the other hand, keyword-based search provides a simple way of retrieving information but is insufficient for identifying meaningful facts and dimensions, as the search results lack context information and do not capture the connections between data elements.

In this paper, we present *SEDA* [1], a system based on a paradigm of search and user interaction. *SEDA* users start with simple keyword style querying, interactively refine the query based on result summaries, and obtain result data cubes that can be further analyzed by off-the-shelf OLAP tools. We envision *SEDA* to be complementary to existing OLAP, data modeling, and ETL tools. The main goal of *SEDA* is to enable users to explore an XML data set, discover interesting dimensions and facts, and pre-analyze the data before creating a full-fledged data warehouse. *SEDA* relies on the collective knowledge of users, who collaboratively define the dimensions and facts of a warehouse schema. This approach is similar to the pay-as-you-go [14] paradigm in that the warehouse schema is defined gradually.

Throughout this paper, we will use an example based on a data scenario constructed from real world data sources. Our example scenario is derived by combining the six annual releases of World Factbook[2] (from 2002 to 2007) and the Mondial XML[3] data. The World Factbook is a publicly available database created and maintained by the Central Intelligence Agency (CIA). It contains comprehensive statistics for every country and territory in the world for each year. The "schema" of this data evolves from year to year. The Mondial data set is a rich compilation of geographical Web data sources on global statistics of world countries, cities, provinces, seas and international organizations. Example data graphs from this data set are shown in Figures 1 and 2, where the solid lines represent parent/child edges and the dashed lines represent non-tree edges[4] denoting various relationships.

---

[*]The work was done while the author was at IBM Almaden Research Center
[1]SEDA stands for Search, Explore, Discover and Analyze

[2]https://www.cia.gov/library/publications/the-world-factbook/
[3]http://www.dbis.informatik.uni-goettingen.de/Mondial/

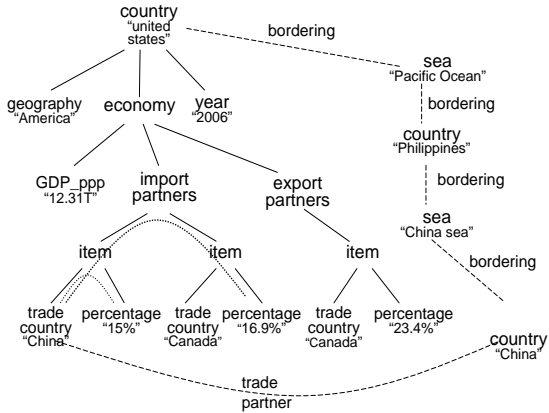

**Figure 1: Example graph from World Factbook & Mondial**

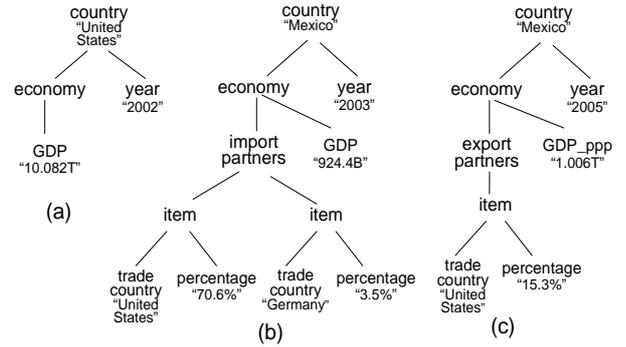

**Figure 2: Example data fragments from World Factbook**

Consider a scenario where a user is interested in finding facts about various countries and continents using this data set, but she is not fully aware of the schema of the data. We will use the following example to illustrate how the user can search, discover and analyze this data set:

EXAMPLE 1. *Consider a user looking for information on import partners of "United States" and their trade amounts. The user starts by searching for "United States" and "import", finds XML fragments similar to Figure 2 (b), realizes that "percentage" information is available, and formulates* **Query 1:** *as follows* $(*, "United States") \wedge (trade\ country, *) \wedge (percentage, *)$

The query language of *SEDA* will be explained in Section 3. For now, it suffices to know that the first component of the query term denotes the query context and is matched to node names and paths, and the second component denotes the content and is matched to the textual content of nodes. The text "United States" occurs in three different contexts in Figures 2 (a), (b), and (c) respectively: in the content of nodes labeled "country" and within nodes labeled "trade country", as an import or an export partner. Each of these three contexts has different real-world semantics and it is very hard for a system to automatically discover what the users' intentions really are. Similarly, trade country and percentage occur in two different contexts, within the context of an import or an export partner. This suggests 12 different ways of combining these nodes.

*SEDA* first quickly retrieves top-k tuples of nodes ranked using a scoring function which considers both the content and the structural connections between the nodes. The score function is based on the compactness of the graph representing a tuple of nodes satisfying query terms. Although *SEDA* employs one particular top-k algorithm, we can also employ any other top-k algorithm that works on graphs.

In addition to the top-k results, *SEDA* also computes two result summaries, called *context* and *connection*, that enable the user to restrict entities and relationships of the query. In our example, the user will be presented with three different contexts for "United States", two different contexts for "trade country" and two different contexts for "percentage". Since the user is interested in import partners of "United States", she can select the corresponding contexts for each term, and refine her query to restrict the results only to those contexts. Even when the user restricts her query to "import partners", there are still two different ways to connect *trade country* and *percentage* elements, represented as dotted lines in Figure 1. *SEDA* computes a set of "meaningful" connections from the top-k results by using dataguide summaries and allows the user to choose which of those are relevant for her query.

Once the user specifies all the contexts and connections that are relevant to her query, she has the option of using the results of this exploration to create an OLAP cube. In this case, *SEDA* computes a full (i.e. not top-k) set of result tuples, with two columns for each query term: The first one contains the Dewey ID [19] XML node reference, and the other one contains the full root-to-leaf path of the node. Figure 3(a) shows two result tuples extracted from the document in Figure 1. Next, *SEDA* automatically matches each path column of this full query result to the contexts of known facts and dimensions. Given a match, it extends the results with necessary key columns, so that we can compute meaningful aggregates. The user also has the option of defining new dimensions and facts based on the columns of their query results. In that case, the system also asks for a key and automatically verifies its uniqueness. Finally, *SEDA* generates the corresponding fact and dimension tables to be fed into an OLAP tool. In our example, assuming that the facts and dimensions of Figure 3(b) already exist in the system, *SEDA* identifies that the query results contain the *country* and *import country* dimensions, as well as the *import trade percentate* fact. If the user agrees with this decision, the system will automatically add the $/country/year$ column to the result, since the key for the *country* dimension[5] contains $/country/year$ path. Moreover, the system will match the newly added $/country/year$ column to the context of the existing *year* dimension, and add this dimension to the output. The final dimension and fact tables generated by *SEDA* for **Query 1** are shown in Figure 3(c). Note, that without the *year* dimension, the fact table would not have a primary key, preventing users from computing meaningful aggregates on this cube. For example, there would be no information on what distinguishes the records that contain "China 12.5%" and "China 13.8%".

We chose to dynamically construct the cube by identifying dimensions and facts in a query result that relevant to a particular analytical task in order to deal with highly heterogenous data sets. For example, query term $(*, "UnitedStates")$ actually matches not 3, but 27 paths in our World Factbook dataset. Some of the matches are in prominent paths, such as /*country*, which occurs in 1577 out of 1600 documents. However, some other paths are more rare such as
/*transnational_issues*/*refugees*/*country_of_origin*,
occurring in only 186 documents. We observe a long tail of such infrequent paths, which makes shredding all the attributes into a data warehouse very difficult.

---
[4]The labels on the dashed lines represent the relationship between the nodes connected by the edge.

[5]In this example all dimensions and facts contain "/country/year" in their key, but it is sufficient for any one of them to have it.

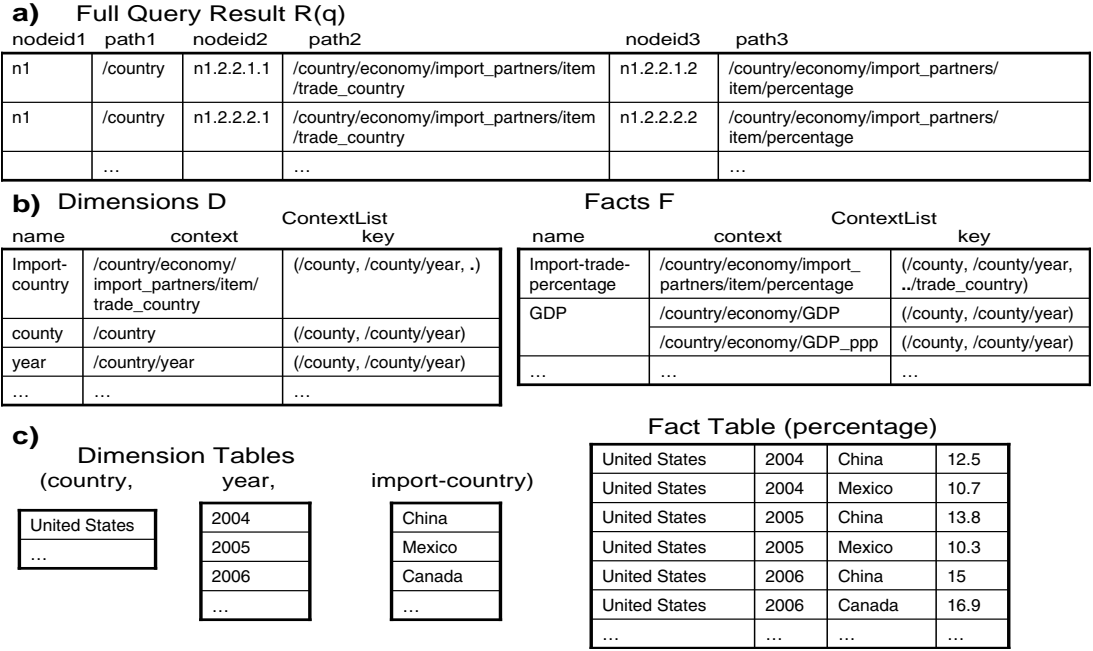

**Figure 3: Sample result for Query 1**

Our main contributions can be summarized as follows:

1. An end-to-end tool to explore and analyze an XML data set to discover interesting properties, facts and dimensions, enabling the creation of a full-fledged OLAP solution.

2. A systematic approach for performing complex analysis of XML data starting with simple keyword-based query terms.

3. Techniques for deriving and representing context and connection summaries to enable user-assisted disambiguation of context and connections, which in turn results in a simplified specification of a precise complex analytic query.

4. Techniques for automatic computation of data cubes from each query result for OLAP-style analysis of the data.

The rest of this paper is organized as follows: We first review related work in Section 2. Section 3 provides some basic definitions. The overview of *SEDA* is in given Section 4. Section 5 and Section 6 describe how context and connection summaries are generated and used in user assisted disambiguation of relevant results, and Section 7 shows how the results of the previous exploration can be used in completing the specification and computation of an OLAP-style aggregation query. Finally, we present our conclusions and some ideas for future work in Section 8.

## 2. RELATED WORK

**XML Flexible Querying:** A number of different approaches have been proposed for flexible querying of XML data[6, 12, 26, 13, 20, 18, 16, 10]. They try to infer and identify meaningfully related answers that are relevant to the user query based on various proposed heuristics. As described in the paper [22], the proposed heuristics do not work on all data scenarios. In *SEDA*, we use a heuristic based on compactness of data graphs augmented with user feedback to disambiguate relevant results. *SEDA* does not limit the connection relationships to tree patterns. It incorporates user feedback, and enables OLAP-style analysis on query results.

**XML Analytics:** A complementary body of work studies grouping and analytics in XML. [3] proposes a group-by operator as an extension to XQuery[25] and presents how complex analysis of XML data can be achieved using the proposed extension. OLAP on XML data incorporates XML grouping techniques to define and compute a data cube such as in [23, 11, 24]. In [24] the data cube is computed based on matching tree patterns with structural relaxations to accommodate for XML's flexible representation. These methods still require the formulation of XQuery queries to perform OLAP computation. Our approach combines search and user guidance to alleviate this issue.

**Faceted Search:** Faceted search engines[2, 21, 7] address the problem of disambiguating multiple real-world entities matching a given keyword by organizing the data into facets and providing a navigational interface. In these systems, the facets are precomputed and the data is indexed accordingly [2, 21, 7]. However, the query language of *SEDA* is more than just simple keywords and as a result, the set of contexts each query term can identify is very large. For example, for the World Factbook data set, we have 1984 distinct paths. As a result, in *SEDA* we have chosen to generate these contexts dynamically for each query.

We adopt a pay-as-you-go approach to defining the data cubes from heterogenous data, which is advocated by Dataspaces [14] data integration work. Unlike Dataspaces which gradually integrates myriad of data sources, the goal of *SEDA* is to identify the part of the data set that is of interest for an OLAP solution, and gradually build up a star schema. In addition, *SEDA* needs to deal with massive variations in data structures due to schema evolution and optional attributes in XML schemas.

## 3. DATA MODEL AND QUERY LANGUAGE

Before describing the details of the *SEDA* system, we first present the underlying data model and the query language of *SEDA*.

*SEDA* operates on a collection of XML documents, which may have links between them. So, we model XML data as a directed graph in which nodes represent element or attribute nodes (referred to as data nodes from now on), and edges represent various relationships between nodes. In this work, we consider four types of relationships between data nodes: (1) parent/child relationship[6], (2) IDREF links, (3) XLink/XPointer links, and (4) valued-based relationships (such as primary key-foreign key relationship). These four relationships convey various semantic information between the participating nodes. We assume that instances of the last type of relationship, i.e. value-based relationships, are provided as input into the system. If not, such relationships can be discovered by employing algorithms to discover keys, such as [27, 17]. Note that discovering and adding appropriate edges into the data graph may require prepocessing of the XML data. We now formally define the data operated on by *SEDA* as follows:

DEFINITION 2 (DATA GRAPH). *A data node $v_1$ is related to another data node $v_2$, if one of the following holds:*

1. $v_1$ *is a child of* $v_2$, *or* $v_2$ *is a child of* $v_1$.
2. $v_1$ *contains an IDREF attribute whose value is equal to the ID attribute of* $v_2$.
3. $v_1$ *contains an XPointer/XLink pointer attribute, whose target node is* $v_2$.
4. $v_1$ *and* $v_2$ *have the same data value, where* $v_1$ *is a primary key and* $v_2$ *is a foreign key.*

*The data graph $G(V, E)$ of an XML collection $C$ is a directed graph where, $V$ is the set of XML element and attribute nodes in $C$, and $E = \{(v_1, v_2) \mid v_1$ is related to $v_2\}$. In G, we distinguish a subset of nodes $Root \subset V$, called root nodes, which are the root elements of the XML documents in collection $C$.*

Each data node has a context and content. We define the *context* of a data node $n$, denoted by $context(n)$, as its root-to-leaf path, starting from a root node $r \in Root$, and following only parent/child edges. We define the *content* of a data node $n$, denoted by $content(n)$, as the concatenation of all the text node descendants of $n$ by traversing parent/child edges only.

The query language of *SEDA* incorporates full-text search as the core component. A query in *SEDA* consists of a set of *query terms*.

DEFINITION 3 (QUERY TERM). *A query term $qt$ is a pair of the form $qt = (context, search\_query)$, where $search\_query$ is any full-text search expression, and $context$ is either empty or one*

---

[6] In this work, we treat element-attribute relationships as a special case of parent/child relationship.

*of the following: (i) a root-to-leaf path, (ii) keyword query, allowing wildcards, or (iii) disjunctions of (i) and (ii).*

*A data node $n$ satisfies a query term $qt$ iff*

1. $Content(n)$ *satisfies* $search\_query$, *and*
2. *One of the following holds:*
    *(1) $qt.context = empty$, or*
    *(2) $qt.context = node\text{-}name(n)$, or*
    *(3) $qt.context = context(n)$, or*
    *(4) $qt.context$ is a disjunction and $\exists\ disjunct\ d \in qt.context$ such that $d = node\text{-}name(n)$ or $d = context(n)$.*

The $search\_query$ of a query term can be a simple bag of keywords, a phrase query or a boolean combination of those. For example, the query term ($country$, "$Romania$") asks for a node named $country$ that contains the keyword "Romania". In its simplest form, a *SEDA* query is a set of query terms. The result of a *SEDA* query is a set of tuples satisfying the query terms.

DEFINITION 4 (QUERY RESULT). *Suppose that a query q contains a set of query terms $QT = \{qt_1, qt_2, \ldots, qt_m\}$. Then, we define the result of q, denoted by $R(q)$, as a set of m-tuples of the form $< n_1, n_2, \ldots, n_m >$, where each $n_i$ satisfies $qt_i$ for $1 \le i \le m$ and there exists a connected data graph $g(V', E') \in G$ with $V' = \{n_1, n_2, \ldots, n_m\}$.*

## 4. SYSTEM OVERVIEW

The architecture of *SEDA* is provided in Figure 4. It contains three major components: a user interface, an execution engine and a storage and indexing component. A user interacts with the system through various panels in the GUI. To help users in the formulation of complex queries, *SEDA* provides context and connection summary panels, as well as an OLAP panel, in addition to the query and result panels. The execution engine contains several units for processing a user's query, guiding her to refine her search for increasing precision and finally for computing OLAP-style data cubes. XML data is stored in DB2® pureXML™[7]. In addition, the storage component contains several indexes to efficiently support these operations, including a Lucene full-text index.

An example screen shot of *SEDA* is given in Figure 5 [1]. In the top panel on the left, the user formulates and refines her query. In this example, the user asks for trade partners of "United States". The control flow of *SEDA* is depicted in Figure 6. When a user submits a query to *SEDA*, it first employs a top-k search unit to compute the most relevant top-k answers fast. In *SEDA*, we employ a top-k search algorithm based on the family of threshold algorithms (TA)[8]. The *SEDA* top-k algorithm retrieves the results from full-text indexes and calculates top answers according to a ranking function which takes into account both the content score as well as the structural properties of the matched nodes. However, we can use any top-k search algorithm that works on data graphs. The results of the top-k search algorithm are shown on the right in the *SEDA* GUI as depicted in Figure 5.

In addition to top-k results, *SEDA* also computes two result summaries, called *context* and *connection*, which help the user to understand the structural properties of the returned results and allow her to further specify additional conditions to express her intention more accurately. The *context summary* is a list of contexts, represented by root-to-leaf paths, in which the query terms are found. The *connection summary* contains all the paths that connect the nodes in the top-k results. Notice from Definition 4 that the number of results, and, thus, the number of distinct paths that connect them can potentially be infinite on graph data. Thus, we limit the relationship summary construction to top-k results. The middle and the bottom panels on the left in Figure 5 show the context and connection summaries for the example query. We describe how *SEDA* computes context and connection summaries in Sections 5 and 6, respectively.

---

[7] DB2 pureXML is a trademark of the IBM Corporation

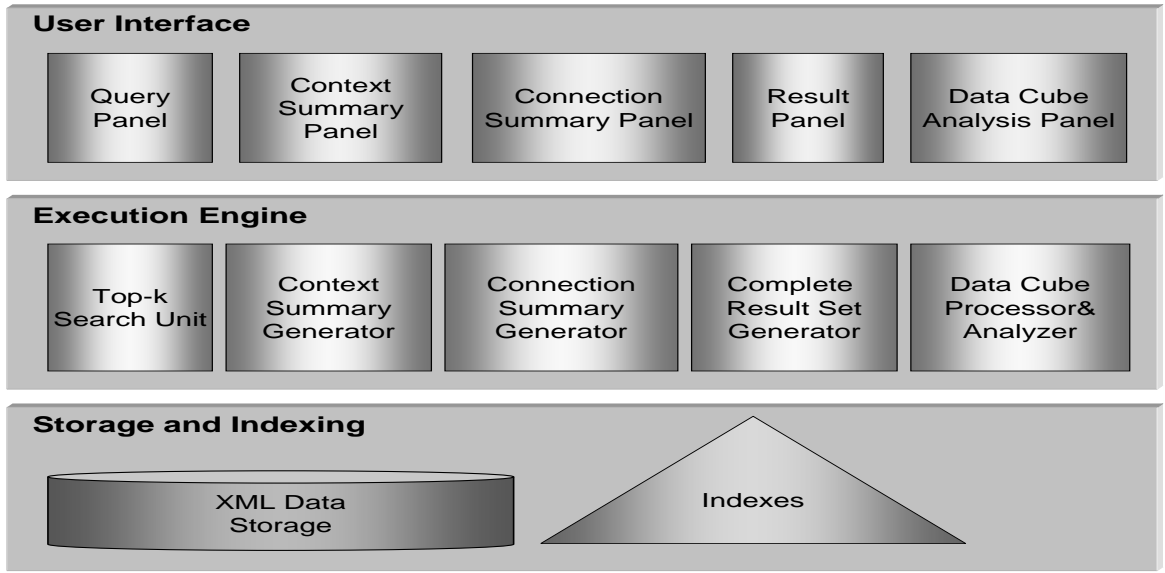

**Figure 4: Architecture of *SEDA***

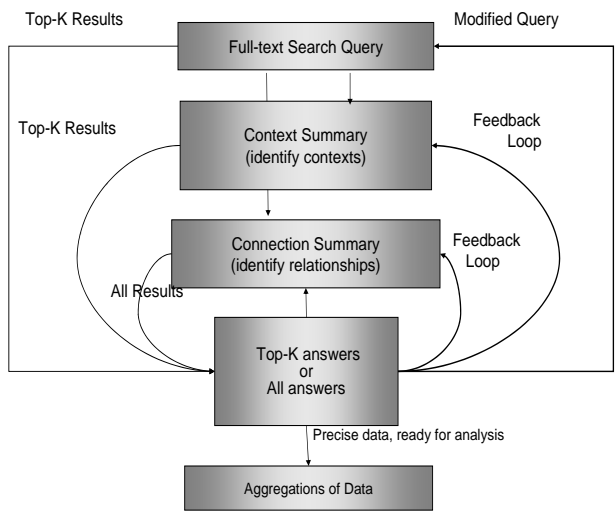

**Figure 6: Control Flow in *SEDA***

If the user finds what she is looking for in the initial top-k results, she may stop the exploration. Otherwise, she may refine her query with the help of the context and connection summaries. If a subset of contexts are chosen, *SEDA* computes the top-k results again limited to this subset. This allows the system to consider only connections between objects relevant to the chosen context set. Once the user selects a subset of connections she is interested in, *SEDA* has sufficient information about the user's intention and now can compute the *entire* result set, not just the top-k, and generate data cubes from the complete results. The rationale is that once the user restricts the contexts and the connections, the search query is refined enough that its result has the precision and recall required for formulating an OLAP-style complex query. To compute the cube, *SEDA* automatically maps each column in the result to a dimension or a fact, and allows the user to choose a subset of dimensions or facts from the set it identified, as well as add other dimensions and facts. For this purpose, *SEDA* provides a data cube screen panel, instead of the top-k results panel, on the left, as shown in Figure 7. The dimensions and measures that match the user query are highlighted in the pulldown menu, but the user is free to choose any dimension and measure she wants. In the final step, *SEDA* automatically generates database queries to create the tables in the corresponding star schema and feeds the tables into an OLAP tool for further analysis. The details are explained in Section 7.

## 5. CONTEXT DISCOVERY

In this section, we show how users may select one or more contexts for each of the query terms to better reflect their intentions. In *SEDA*, we represent the context of nodes in terms of their root-to-leaf paths. Additionally, a context can be further abstracted and represented by a real-world entity, if such information is available. Consider **Query 1** in the introduction, which asks for the percentage trade amounts of import trade partners of "United States" and the data in Figure 2. The phrase "United States" will be found in three different contexts: as a country name in Figure 2 (a), as an import partner trade country in Figure 2(b) and as an export partner trade country in Figure 2(c). Each of these three contexts has different real-world semantics which makes it challenging for a system to automatically discover what the users' intentions really are.

Search engines address this problem by providing a faceted search interface, so that users can choose the context they are interested in. While faceted search engines precompute facets, in *SEDA* we have chosen to generate these contexts dynamically for each query. But,

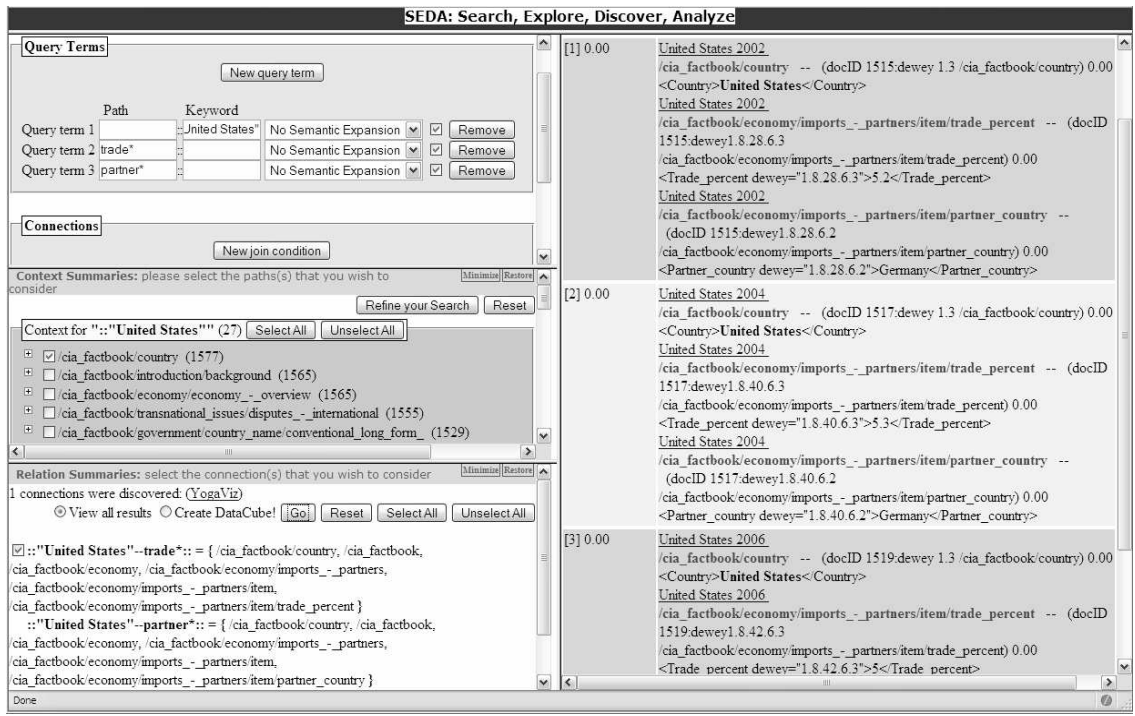

Figure 5: Example screen shot of *SEDA*

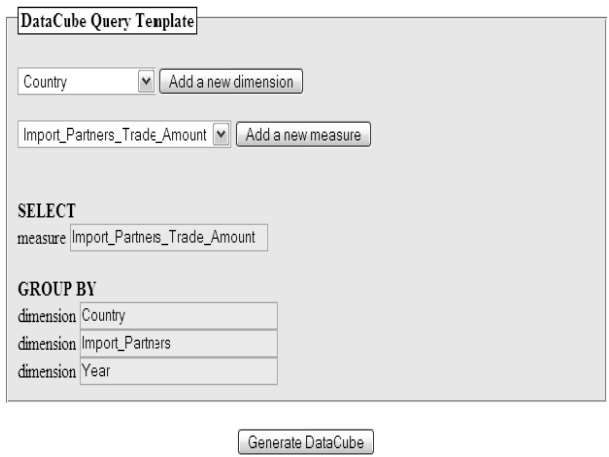

Figure 7: Data cube specification user interface.

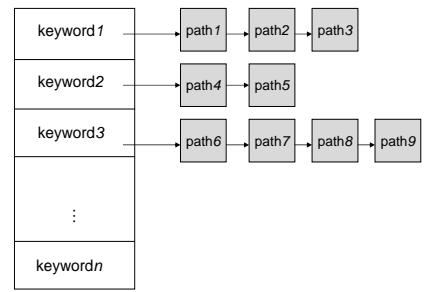

Figure 8: Full-text index for contexts

This full-text index contains all keywords that appear in the data set as content, as well as all the tag names. Each distinct path is treated as a virtual document. Hence, the posting lists contain all the paths a given word appears in. We store the count of occurences of each path in the document store. An alternative design, which would avoid accessing the document store, would be to store the counts for each path in the posting lists. But, then this would have the drawback of repeating the same information in many posting lists.

The exact usage of this index depends on the query term. If a query term contains only a full-text search query, we take that search query and run it on the full-text index directly. If the query term also contains a full root-to-leaf context, then we use the last tag name in the full root-to-leaf path to probe the index, in conjunction with the search query, and compute the frequency of the as in the case of faceted search, we still rely on the user to choose the appropriate context.

Given a *SEDA* query $q$, for each query term, we compute a context bucket that contains all distinct paths that the query term appears in within the entire data collection. To compute this efficiently, we maintain a full-text index which maps individual keywords to the set of distinct paths in which they appear. This allows us to compute the set of distinct paths for phrase queries, as well as other search queries with multiple keywords connected with conjunction or disjunction. This index is illustrated in Figure 8.

path. If the context of the query term is only a tag name, including wildcards, then we use the tag name, in conjunction with the search query to probe the index.

*SEDA* displays all contexts for a query term sorted by their frequencies. *SEDA* shows the frequencies in the entire data collection, not the frequencies in the query result. Note that this is different than faceted search engines which show the frequency of a given keyword in the corresponding path. Instead we display the absolute frequency of the path itself, irrespective of the keyword. The rationale behind this choice is to give the user some idea about the structural properties of the data. Once the user specifies the set of contexts she is interested in, *SEDA* re-computes top-k results, with the additional constraint that the results satisfy the contexts chosen by the user.

## 6. RELATIONSHIP DISCOVERY

Once the user restricts the set of contexts she is interested in, the set of alternative connections *SEDA* needs to consider is reduced. However, there may still be some ambiguity. Recall from **Query 1** that even when the user chooses one context for each query term, restricting the results to only import partners, there are still two different ways to connect *trade country* and *percentage* data nodes, shown as dotted lines in Figure 1. This case is an example when there are different paths between different instances of two types. In another scenario, there maybe multiple paths connecting two data node instances, each representing a different real-world relationship, in a data graph. For example, in Figure 1, we see two paths between "United States" and "China", one representing the geographical relationship and the other one representing the trade partnership.

There are various heuristics proposals in the literature to decide which connections are more meaningful. But as shown in [22], these heuristics work in some scenarios but fail in others. It is very hard to find an approach that will work in all possible data and query scenarios, because each different connection represents a different semantic relationship and it is impossible to know what the user intension really is. In *SEDA*, we rely on user feedback, instead of heuristics, to decide which connections are relevant for the user's query.

A straight-forward algorithm to compute all possible graphs that connects user's contexts is as follows: First compute the complete result set of $q$ on $G$, map the nodes in the result tuples onto $G$, and identify all connected graphs. However, this is a very expensive operation, because the number of possible connections is usually pretty large. It may even be infeasible to show all connections between matching nodes on graph data.

The solution employed by *SEDA* is to choose a subset of "meaningful" connections to present to the user, and let her specify the ones that are relevant for her query. Given the large number of possible connections, the challenge is to discover a subset that is meaningful. We refer to this subset of connections to be presented to the user as the *connection summary*. Our approach is to use the result set, $R^{top-k}(q)$, generated by the top-k algorithm, instead of the complete result set $R(q)$, to extract and create a set of connections to be presented to the user. In addition, instead of computing connected graphs, we show pairwise connections between the matching nodes. Although we are biased towards the top-k results, we expect users to select the contexts first to identify the set of objects they are interested in. Once the user chooses the set of contexts she is intested in, top-k results contain only connections between those relevant contexts and we will be able to capture the meaningful connections.

*SEDA* displays these connections in a visual graph representation and allows the user to pick or drop connections from the connection summary. Once the user selects the set of connections that are relevant for her query, *SEDA* refines the search results to include result tuples that satisfy only those connections.

### 6.1 Computation of Connection Summary

We now describe the algorithm for computing the connections in the connection summary. We assume that the data graph $G$ does not fit in memory, which we expect to be the case for general XML data collections. We observe that we do not necessarily need to work on $G$ itself. Instead, we can compute and use a *summary* of the structural organization of $G$.

For the implementation of the algorithm, we decided to leverage the dataguide structures [15, 9] over XML data. We first compute a collection of dataguides, denoted by $DG$, together with a set of links between the dataguides corresponding to the external edges between documents in $G$ (e.g., IDREF links, XLink/XPointer links and value-based links). We represent a dataguide $dg$ as a list of full root-to-leaf paths such that every full root-to-leaf path in $G$ maps onto a full root-to-leaf path in one $dg \in DG$.

The dataguide computation algorithm computes the summary of each document, i.e. its dataguide, and tries to merge this dataguide with existing dataguides that have been computed so far. As a result, it has a computational cost of $O(n*m)$, where $n$ is the number of documents, and $m$ is the number of dataguides. When we try to merge the dataguide of an individual document, say $dg(d)$, with existing dataguides, we have three possibilities. $dg(d)$ might be a subset of or be exactly the same as one of the existing dataguides, or it may overlap with some of them. In the first two cases, we do not need to do any further processing. In the final case, we have the option of generating a separate dataguide. But, this may not summarize the data as much as we want. For example, we observed that on our World Factbook data set, we created 1600 dataguides for 1600 XML documents due to a large number of optional elements. The other alternative is to merge the new dataguide $dg(d)$ with an existing dataguide based on some similarity measure, which reduces the number and the total size of dataguides in the summary. For this purpose, we use a similarity metric based on the overlap between the set of paths belonging to two different dataguides, given by:

$$overlap(dg_1, dg_2) = min \left\{ \begin{array}{c} \frac{|common\_paths(dg_1, dg_2)|}{|paths(dg_1)|}, \\ \\ \frac{|common\_paths(dg_1, dg_2)|}{|paths(dg_2)|} \end{array} \right\}$$

where, $common\_paths(dg_1, dg_2)$ denotes the set of common root-to-leaf paths between $dg_1$ and $dg_2$, and $paths(dg)$ denotes the set of root-to-leaf paths in $dg$. We merge two dataguides if their overlap factor is over a given threshold. We ran our algorithm on different data sets and observed that the effectiveness of the overlap threshold in reducing the total number of generated dataguides depends on the dataset, ranging from a factor of 3 to a factor of 100 reduction. Table 1 shows the dataguide statistics for an overlapping threshold of $40\%$ for four different data sets that we examined. We observe that the effectiveness of the overlap threshold in reducing the total number of generated dataguides depends on the dataset. For example, for datasets, such as the Google Base, where the data schema is flat and regular, we observe a reduction of upto two orders of magnitude. At the other end of the spectrum, for very flexible datasets, such as for the World Factbook, we see a reduction factor of only 3.

It is important to note that there may be connections discovered from the dataguides, which do not have any instantiation in the portion of data graph $G$ induced by the query result. We refer to these cases as *false positives*. There are two reasons: First, the actual hits in the query result are further restricted by keyword search terms. However, the dataguide contains all connections, irrespective of content values. Second, merging similar dataguides introduces some false connections. Hence the higher the overlap threshold, the fewer the false positive connections because there will be fewer dataguide merges.

Table 1: Dataguide statistics for threshold of 40%

| Data set | # documents | # data guides |
|---|---|---|
| Google Base snapshot [8] | 10000 | 88 |
| Mondial [9] | 5563 | 86 |
| RecipeML [10] | 10988 | 3 |
| World Factbook 2007 [11] | 1600 | 500 |

The dataguide summary is precomputed on the entire data graph $G$. At query time, *SEDA* optimizes the use of the dataguide index by loading it into memory only once from disk. Each time *SEDA* runs the top-k algorithm, it also passes the result set to the connection summary computation algorithm, which maps nodes from the top-k result $R^{top-k}(q)$ onto nodes in the dataguide set $DG$ based on matching root-to-leaf paths. If there are multiple paths between two dataguide nodes, the algorithm chooses the one with the shortest path in $DG$. As an optimization, we cache the connections we discover so that we can leverage the cache for later query hits. Finally, we compute the connections between the matching dataguide nodes.

## 7. CONSTRUCTING DATA CUBES

We facilitate further analysis of the query results by deriving and populating a star schema that is directly usable by an off-the-shelf OLAP engine. But first, we need to materialize the complete set of results for a user query, not just the top-k results. For each connection chosen by the user, the *complete result generator* in Figure 4 computes the complete result set that satisfies the query and the connection constraints.

For each connection chosen by the user, the nodes and all connections together form a connection graph. We partition each connection graph into *twigs*. Each twig is a query pattern tree [4], which includes the connection nodes and parent/child edges within the same document. The remaining edges are called *cross-twig* joins, which combine the results from different twigs. The input to a leaf node in a twig are generated from its full-text search. We retrieve the data nodes from the full-text search results in Dewey ID [19] order, which can be directly used by the XML twig processing [4]. After we compute the results of each twig query, we *join* the results from different twigs according to the *cross-twig* join edges to produce the complete result tuples, which is similar to a join in an RDBMS.

Once the system computes the complete result set of a query, the user has an option to compute one or more data cubes from the results for further analysis. *SEDA* automatically maps the components of the result tuples onto the set of known dimensions and facts, and, after optional manual adjustments to the mapping, computes an instantiation of the corresponding star schema.

To facilitate cube computation, *SEDA* maintains a set of facts, denoted by $F$, and a set of dimensions, denoted by $D$, known to the system. These sets are initially provided by a system administrator and are expanded by users during query processing, i.e. any user can define new facts and dimensions based on their query results. *SEDA* could also take advantage of automated discovery of facts and dimensions.

*SEDA* requires every dimension table to have a key in order to have meaningful aggregates. We employ *relative* XML keys [5]. A relative key for an XML node $n$ is defined as a list of paths $(P_1, P2, \ldots, P_m)$, where each $P_i, 1 \leq i \leq m$, is either an absolute path expression, which starts at the root of the document, or a relative path expression, which starts at the node $n$. If we consider the *percentage* child of *import partners* in Figure 1, we see that there are multiple instances of the *percentage* element, and we need to make sure that we pair China with 15% and Canada with 16.9%. Thus, the key for the *percentage* fact is $(/country, /country/year, ../trade\_country)$. The first two components of this key are absolute and last one is relative to the $/country/economy/import\_partners/item/percentage$ context. I.e. for every *percentage* the key contains its $trade\_country$ sibling. This assumes that every *percentage* in the result will have exactly one such sibling, as well as that every document in the result will have exactly one $/country$ and $/country/year$ elements.

The set of facts, $F$ is defined as a nested relation, with the following schema: $< name, ContextList >$, where $name$ is the name of the fact, and the $ContextList$ is another relation and has the schema $< context, key >$. The first component of the $ContextList$ relation identifies the context of the fact (i.e. root-to-leaf path), and the second component is the relative key w.r.t. the node identified by the context. The reason why $ContextList$ is a relation is because the underlying data collection may be heterogeneous, with documents having no schema or conforming to many different schemas. This heterogeneity may arise from schema evolution or from data integration if the data was collected from many different sources. For the data in Figure 2, the GDP fact is defined by two paths, "$/country/economy/GDP$" and "$/country /economy/GDP\_ppp$". In the case of World Factbook data, the schema has evolved over time and documents created before 2005 have the GDP value in an element called "$GDP$", whereas documents created after 2005 have the corresponding value in an element called "$GDP\_ppp$". The set of dimensions, $D$, is defined similarly. Figure 3 shows an instance of $F$ and $D$ for the World Factbook data set. It is important to note that $F$ and $D$ contain path information defining the facts and dimensions, and do not contain any instance values.

We are now ready to describe how *SEDA* computes the fact and dimension tables from a query result. There are three major steps: (1) matching the query results to facts and dimensions, (2) augmenting the result set, and (3) generation of SQL/XML queries to extract the values from the database to compute the fact and dimension tables. In the following, we will denote a single fact in $F$ with $f$, and a single dimension in $D$ with $d$.

**Step 1: Matching:** The input to this step is the query result $R(q)$ and the sets $F$ and $D$. Recall that the result $R(q)$ of a user query $q$ is a set of tuples. Let $< c_{n1}, c_{p1}, c_{n2}, c_{p2}, \ldots, c_{nm}, c_{pm} >$ denote the schema of $R(q)$, where each $c_{ni}$ represents the dewey ids of the nodes that satisfy a query term $qt_i$, and $c_{pi}$ denotes the contexts of those nodes. Figure 3 shows an instance of $R(q)$ for **Query 1**.

We say that a pair $(c_{ni}, c_{pi})$ matches a fact $f$ iff $\pi_{c_{pi}}(R) \subseteq \pi_{context}(f.ContextList)$. In other words, a pair $(c_{ni}, c_{pi})$ matches a fact $f$ if the set of paths that satisfy the query term $qt_i$ is a subset of the paths that define $f$. A dimension match is defined similarly. In addition to a complete match, there are two other possible scenarios when matching query results to facts and dimensions. In the first case, a path column of the result set $R(q)$ does not intersect with the context list of any known fact or dimension. In this case, the user has the option of defining a new dimension or a fact from that column. The new facts and dimensions can be created in

*SEDA* by specifying a unique name, whether it is a fact or a dimension, and a key for each path. The system automatically verifies the keys by computing them for every $c_{ni}$ in $R(q)$ and checking their uniqueness. Currently, the keys are specified manually, but in the future we plan to adopt the techniques of GORDIAN [17] to discover them automatically. In the second case, some but not all paths in a path column of the result set $R(q)$ intersect with the contexts of a fact or a dimension. In this case, *SEDA* issues a warning message to the user to make sure that she has chosen the correct context list. Once again, the user has an option of defining this column as a new dimension or fact. In both cases, if the user does not create a new fact or dimension for this column, we simply ignore it while creating the cube. The rationale is that those values may have been used only to restrict the data set and are not needed in the data cube computation.

For **Query 1**, $R(q)$ in Figure 3, the first two columns will match the *country* dimension, the third and fourth columns will match the *trade country* dimension, and finally the last two columns will match the *percentage* fact. As a result of the matching process, *SEDA* identifies $F_q$ and $D_q$, the set of facts and dimensions in the result set of $q$, respectively.

**Step 2: Augmentation:** The purpose of this step is to manually augment the automatically matched sets of facts and dimensions, if necessary. *SEDA* allows users to add and remove facts and dimensions to $F_q$ and $D_q$ sets. The result of this step are final sets of facts and dimensions: $F_{final}$ and $D_{final}$.

Given the sets $F_{final}$ and $D_{final}$, we may need to expand $R(q)$ to make sure it includes all key and value columns of every fact and dimension. Consider $R(q)$ in Figure 3, and the *percentage* fact. The relative key of *percentage* consists of $(country, year, trade\ country)$, and $year$ is not in $R(q)$. Without the year values, we cannot create a well-defined fact table and compute meaningful aggregates.

**Step 3: Extraction:** Once we compute the sets $F_{final}$ and $D_{final}$, we generate database queries to compute the fact and dimension tables in the corresponding star schema. Note that *SEDA* stores XML data in a database, and the complete result set $R(q)$ contains only the node-ids for the matching nodes. To compute fact and dimension tables, we need to get their corresponding data values. For those facts (dimensions) in the query result, i.e. $f \in F_q$ ($d \in D_q$), we only need to access the database to retrieve the content of the nodes, whose node-ids are in $R(q)$. For the extra facts (dimensions) the user has added, i.e. $f \in F_{final} \wedge f \notin F_q (d \in D_{final} \wedge d \notin D_q)$, we also need to access the XML document to first locate the correct node and then retrieve its content. In generating those database queries, we use the keys to make sure we pair up correct values.

We generate one fact table for each fact $f \in F_{final}$ and create one dimension table for each dimension $d \in D_{final}$. For example, for the *percentage* fact of **Query 1**, we generate the fact table in Figure 3. Note that all the components of the key of the *percentage* fact are in the fact table, including the additional *year* column. As an optimization, we merge fact tables if they have the same keys. Therefore, the outcome of this phase is a set of fact and dimension tables. It is also important to note that each user query may identify different sets of facts and dimensions, resulting in a different star schema. This is in contrast with traditional data warehouses, where there is one fixed star schema. This dynamic computation of a different star schema instance for each user query allows the users to analyze their data on-demand.

Finally, we feed these tables into an OLAP-tool to compute the data cubes, one per fact table, and the desired aggregation functions for further analysis.

## 8. CONCLUSION AND FUTURE WORK

In this paper, we described a prototype system, *SEDA*, which is based on a paradigm of search and user interactions to help users start with simple keyword style querying and perform rich analysis of XML data. We envision it to be used as a tool to explore and analyze an XML data set to discover interesting properties, facts and dimensions, enabling the creation of a full-fledged OLAP solution. By dynamically computing a star schema and its instantiation for each query, *SEDA* allows on-demand analysis of XML data.

As future work, we plan to investigate automatic discovery of facts and dimensions from the data, as well as improvements to our summarization algorithms. Another interesting avenue of research is defining proper metrics to evaluate a system like *SEDA* in terms of its effectiveness.

## 9. ACKNOWLEDGEMENTS

We would like to thank Zografoula Vagena for her contributions to the top-k search algorithm and some other aspects of the system.

## 10. REFERENCES

[1] A. Balmin, L. Colby, E. Curtmola, Q. Li, F. Özcan, S. Srinivas, and Z. Vagena. SEDA: A System for Search, Exploration, Discovery, and Analysis of XML Data. In *Proc. of VLDB*, pages 1408–1411, 2008. Demo paper.

[2] Ben-Yizhak et al. Beyond Basic Faceted Search. In *Proc. of WWW*, 2007.

[3] K. Beyer, D. Chamberlin, L. Colby, F. Özcan, H. Pirahesh, and Y. Xu. Extending xquery for analytics. In *Proc. of SIGMOD*, pages 503–514, 2005.

[4] N. Bruno, N. Koudas, and D. Srivastava. Holistic twig joins: optimal XML pattern matching. In *Proc. of SIGMOD*, pages 310–321, 2002.

[5] P. Buneman, Susan B. Davidson, W. Fan, Carmem S. Hara, and Wang Chiew Tan. Keys for xml. In *Proc. of WWW*, pages 201–210, 2001.

[6] S. Cohen, J. Mamou, Y. Kanza, and Y. Sagiv. XSEarch: A Semantic Search Engine for XML. In *Proc. of VLDB*, pages 45–56, Berlin, Germany, 2003.

[7] W. Dakka, R. Dayal, and P. Ipeirotis. Automatic Discovery of Useful Facet Terms. In *Proc. of SIGIR Workshop on Faceted Search*, pages 768–775, 2006.

[8] R. Fagin, A. Lotem, and M. Naor. Optimal Aggregation Algorithms for Middleware. In *Proc. of PODS*, 2001.

[9] R. Goldman and J. Widom. Dataguides: Enabling query formulation and optimization in semistructured databases. In *Proc. of VLDB*, pages 436–445, 1997.

[10] L. Guo, F. Shao, C. Botev, and J. Shanmugasundaram. XRANK: ranked keyword search over XML documents. In *Proc. of SIGMOD*, pages 16–27, San Diego, USA, 2003.

[11] C. A. Hurtado and A. O. Mendelzon. Reasoning about summarizability in heterogeneous multidimensional schemas. *Lecture Notes in Computer Science*, 1973, 2001.

[12] Y. Li, C. Yu, and H. Jagadish. Schema-Free XQuery. In *Proc. of VLDB*, pages 72–83, Toronto, Canada, 2004.

[13] Z. Liu and Y. Chen. Identifying Meaningful Return Information for XML Keyword Search. In *Proc. of SIGMOD*, Beijing, China, 2007.

[14] J. Madhavan et al. Web-Scale Data Integration: You can afford to Pay as You Go. In *Proc. of CIDR*, 2007.

[15] S. Nestorov, J. Ullman, J. Wiener, and S. Chawatbe. Representative Objects: Concise Representations of Semistructured Hierarchical Data. In *Proc. of ICDE*, 1997.

[16] T. Saito and S. Morishita. Amoeba Join: Overcoming Structural Fluctuations in XML Data. In *Proc. of WebDB*, pages 38–43, Chicago, USA, 2006.

[17] Y. Sismanis, P. Brown, Peter J. Haas, and B. Reinwald. GORDIAN: Efficient and Scalable Discovery of Composite Keys. In *Proc. of VLDB*, pages 691–702, 2006.

[18] C. Sun, C.-Y. Chan, and A. K. Goenka. Multiway SLCA-based Keyword Search in XML Data. In *Proc. of WWW*, Singapore, Singapore, 2007.

[19] I. Tatarinov et al. Storing and Querying Ordered XML Using a Relational Database System. In *Proc. of SIGMOD*, 2002.

[20] M. Theobald, R. Schenkel, and G. Weikum. An Efficient and Versatile Query Engine for TopX Search. In *Proc. of VLDB*, pages 625–636, Trondheim, Norway, 2005.

[21] D. Tunkelang. Dynamic Category Sets: An Approach for Faceted Search. In *Proc. of SIGIR Workshop on Faceted Search*, 2006.

[22] Z. Vagena, L. Colby, F. Özcan, A. Balmin, and Q. Li. On the Effectiveness of Flexible Querying Heuristics for XML Data. In *XSym*, pages 77–91, 2007.

[23] H. Wang, J. Li, Z. He, and H. Gao. OLAP for XML Data. In *CIT*, pages 233–237, 2005.

[24] N. Wiwatwattana, H. Jagadish, L. Lakshmanan, and D. Srivastava. $X^3$: A Cube Operator for XML OLAP. In *Proc. of ICDE*, 2007.

[25] *XQuery 1.0: An XML Query Language*, January 2007. W3C Recommendation, See http://www.w3.org/TR/xquery.

[26] Y. Xu and Y. Papakonstantinou. Efficient Keyword Search for Smallest LCAs in XML Databases. In *Proc. of SIGMOD*, pages 537–538, Baltimore, USA, 2005.

[27] C. Yu and H.V. Jagadish. Efficient Discovery of XML Data Redundancies. In *Proc. of VLDB*, pages 103–114, 2006.